\documentclass[preprints,article,accept,oneauthor,pdftex,10pt,a4paper]{mdpi} 

\firstpage{1} 
\history{Condens. Matter {\bf 3}, 7 (2018) \\
DOI:\href{http://dx.doi.org/10.3390/condmat3010007}{10.3390/condmat3010007}}

\pdfoutput=1

\usepackage{graphicx}
\usepackage{amsmath}

\Title{Ground-State Magnetization in Mixtures of a Few Ultra-Cold Fermions in One-Dimensional Traps}

\Author{Tomasz Sowi\'nski\orcidA{}}

\AuthorNames{Tomasz Sowi\'nski}

\address[1]{Institute of Physics, Polish Academy of Sciences, Aleja Lotnikow 32/46, PL-02668 Warsaw, Poland; tomsow@ifpan.edu.pl}

\abstract{Ground-state properties of a few spin-$1/2$ ultra-cold fermions confined in a one-dimensional trap are studied by the exact diagonalization method. In contrast to previous studies, it is not assumed that the projection of a spin of individual particles is fixed. Therefore, the spin is treated as an additional degree of freedom and the global magnetization of the system is established spontaneously. Depending on the shape of the trap, inter-particle interactions, and an external magnetic field, the phase diagram of the system is determined. It is shown that, for particular confinements, some values of the magnetization cannot be reached by the ground-state of the system.}


\begin{document}

\section{Introduction}
Recent ground-breaking experiments on ultra-cold atoms confined in quasi-one-dimensional optical traps have opened a completely new avenue for studying problems of a few quantum particles~\cite{Paredes2004tonks,Kinoshita2004observation,Guan2013,haller2009realization}. It becomes possible to control not only mutual interactions and the external confinement that the particles are stored in but also to control the number of particles in experiments repeated with tremendous accuracy \cite{Murmann2015AntiferroSpinChain,Kaufman2015Entangling,serwane2011deterministic,wenz2013fewToMany}. In consequence, ultra-cold atomic systems become dedicated simulators for systems with a mesoscopic number of particles, yielding an ability to study different properties of bosonic and fermionic systems and a variety of their mixtures~\cite{zurn2012fermionization,zurn2013Pairing,Murmann2015DoubleWell,Sowinski2015Pairing,DAmico2015Pairing,BjerlinReimann2016Higgs,GarciaMarch2014,GarciaMarch2014Localization,Zinner2014Bosons1D,Harshman2017}.

Typically, in the context of mixtures of a few interacting ultra-cold fermions, it is assumed that particles from different components belong to different, fundamentally distinguishable, families~\cite{Blume2012Rev,Zinner2016Rev}. This assumption is well justified from the experimental point of view since different fermionic components are usually formed from different atomic elements \cite{Blume2008,Mehta2015,Blume2010,Daily2012,Dehkharghani2016,Pecak2016} or from atoms belonging to different irreducible spin representations~\cite{serwane2011deterministic,Sowinski2013,Deuretzbacher2014,Mehta2014,GarciaMarch2014,Grining2015,Hu2016,Harshman2016I,Harshman2016II}. In such a case, particles are simply forbidden to change their flavor, and, as a consequence, the number of particles of a given type is conserved. 

Here, we consider a slightly different experimental situation in which projections of the individual spins are not fixed and may be freely distributed among different spin components, i.e., only the total number of particles is conserved. Consequently, the system possesses an additional degree of freedom that can be directly utilized to avoid interactions and to minimize the energy of the many-body ground-state. The distribution of particles among different species is established spontaneously in response to external parameters being under experimental control. For example, when interactions between particles belonging to different components are very strong, it may be energetically favorable to transfer all particles to a single spin-flavor, since the energy cost needed to excite particles may be smaller than the energy gain from avoiding interactions. This energetic argumentation holds for repulsive as well as for attractive interactions. In the following, we mainly focus on the repulsive branch, since attractive interactions have rather trivial consequences to the magnetization of the ground-state of the system.

To make the analysis as simple as possible, in this paper, we assume that all considered atoms are indistinguishable fermionic spin-$1/2$ particles. Interactions between particles are modeled with a zero-range $\delta$-like potential. In consequence, due to fermionic statistics, mutual interactions are present only between particles of opposite spins. This approximation is well justified experimentally when one considers collisions of ultra-cold atoms \cite{PethickBook}. To emphasize a role of the allowed spin-flipping mechanism, we will consider different arrangements of the external potential. Moreover, we will discuss the role of the external magnetic field, which is known to have a substantial influence on arrangements of atomic magnetic moments (originating in the non-vanishing total spin of atoms) in ultra-cold clouds. Moreover, starting from seminal works \cite{Kawaguchi2006,Santos2006,Gawryluk2007}, it is known that the magnetic field is crucial when the famous Einstein--de Haas effect driven by spin-flipping is considered.

The main goal of this work is to settle how the total magnetization of the many-body ground-state (the sum of all projections of individual spins) depends on the interaction strength, the shape of the confinement, and the value of the external magnetic field. 

\section{The System Studied}

With these motivations, we consider an ultra-cold system of interacting spin-1/2 particles of mass $m$ confined in a one-dimensional trap $V_\xi(x)$ with a shape controlled by one parameter $\xi$. The assumption of a one-dimensionality is well justified in view of present experiments on ultra-cold particles since it is possible to control the shape of an external potential in all three spatial directions independently. By applying very deep confinement in two perpendicular directions, one can force all atoms to occupy the single-particle ground-state in both of these directions. In consequence, all properties of the system are determined by the one-dimensional physics in the remaining direction.

\begin{figure}[!hbtp]
\centering
\includegraphics[scale=1.1]{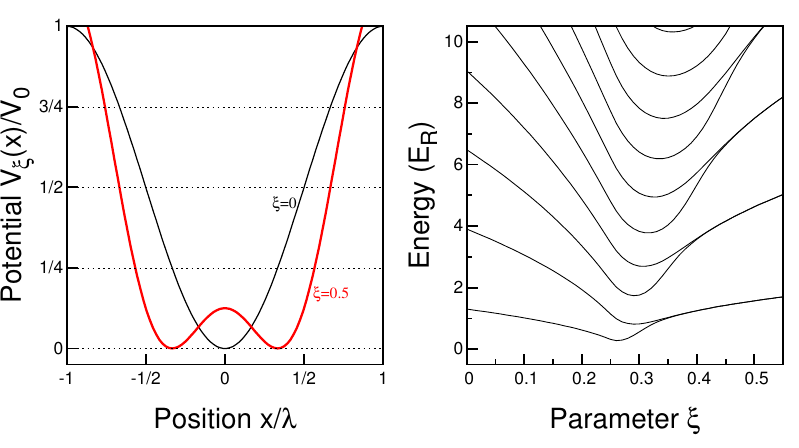}
\caption{Shape of the external potential $V_
\xi(x)$ (\textbf{left}) and the single-particle spectrum of the corresponding Hamiltonian $H_\xi$ (\textbf{right}) as functions of the parameter $\xi$ controlling the barrier between the wells. For a sufficiently deep barrier, the single-particle spectrum becomes doubly quasi-degenerated. \label{Fig1}} 
\end{figure}

In the following, an external potential $V_\xi(x)$ is modeled as a one-dimensional optical superlattice forming a periodic structure of local double-well potentials:

\begin{equation} \label{Potential}
V_\xi(x) = \frac{V_0}{2}\left[1 + \xi - \cos\left(\frac{\pi x}{\lambda}\right) + \xi \cos\left(\frac{2\pi x}{\lambda}\right)\right]-U(\xi)
\end{equation}
where $\lambda$ is the laser wavelength, and $V_0$ is a depth of the optical lattice controlled experimentally~\cite{Valtolina2015,Murmann2015DoubleWell,Tylutki2017}. All physical quantities can be expressed in the natural units of the problem, i.e., the laser wavelength $\lambda$ and the recoil energy $E_R=2\pi^2\hbar^2/m\lambda^2$ are taken as units of length and energy, respectively. We assume that $V_0$ is large enough to neglect all tunnelings between lattice sites (we set $V_0=100E_R$). A relative intensity of the additional optical lattice with half-wavelength is controlled by $\xi$. For $\xi=0$, the potential reproduces almost ideally a harmonic trap with frequency $\Omega=\sqrt{2\pi^2V_0/m\lambda^2}$. When $\xi$ is positive additional repulsion in the middle of the trap arises and starting from $\xi=0.25$, the potential has two symmetrically placed wells (see the left panel in Figure~\ref{Fig1}). For convenience, we introduce also an additional shift of the energy $U(\xi)$, which simply assures that, for any $\xi$, the potential (\ref{Potential}) has minima at zero energy. 
 
Since particles have the internal spin degree of freedom, we assume that they can be subjected to an external magnetic field $B$ in the direction of the spin quantization axes. In this way, the relative single-particle energy gap between components can be controlled. Fermions with a given spin $\sigma\in(\uparrow,\downarrow)$ are described with a corresponding quantum field $\hat{\psi}_\sigma(x)$ obeying standard anti-commutation relations

\begin{equation}
\left\{\hat\psi_\sigma(x),\hat\psi^\dagger_{\sigma'}(x')\right\}=\delta_{\sigma\sigma'}\delta(x-x'), \qquad
\left\{\hat\psi_\sigma(x),\hat\psi_{\sigma'}(x')\right\}=0.
\end{equation}

For further convenience we also introduce a standard two-component spinor field $\hat{\boldsymbol{\Psi}}^\dagger(x)=\left(\hat\psi^\dagger_\uparrow(x),\hat\psi^\dagger_\downarrow(x)\right)$ describing both components simultaneously. 
The many-body Hamiltonian of the system has a form

\begin{equation} \label{Ham}
\hat{\cal H} = \int\!\!\mathrm{d}x\,\,\hat{\boldsymbol{\Psi}}^\dagger(x)\left(H_\xi + B\sigma_z\right)\hat{\boldsymbol{\Psi}}(x) 
+ g\int\!\!\mathrm{d}x\,\,\hat{\psi}_\uparrow^\dagger(x)\hat{\psi}_\downarrow^\dagger(x) \hat{\psi}_\downarrow(x)\hat{\psi}_\uparrow(x)
\end{equation}
where $g$ measures the  strength of the contact interactions. Note that an external magnetic field $B$ is measured in natural units of energy, i.e., the magnetic moment of particles is incorporated to $B$. In a one-dimensional scenario, the interaction strength $g$ can be well controlled not only by an external magnetic field (via Feshbach resonance) but also by the trapping frequency of a perpendicular confinement \cite{Olshanii1998,Bergeman2003}. Therefore, it can be treated as an independent parameter controlled experimentally.  The single-particle part of the Hamiltonian reads

\begin{equation} \label{HamSingle}
H_\xi = -\frac{1}{2}\frac{\mathrm{d}^2}{\mathrm{d}x^2} + V_\xi(x).
\end{equation}

It is quite obvious that the many-body Hamiltonian (\ref{Ham}) commutes with operators of the number of particles $\hat{\cal N}$ and the spin-imbalance $\hat{\cal M}$ defined as

\begin{equation} \label{OpMN}
\hat{\cal N}=\int\mathrm{d}x\,\,\hat{\boldsymbol\Psi}^\dagger(x)\hat{\boldsymbol\Psi}(x), \qquad \hat{\cal M}=\int\mathrm{d}x\,\,\hat{\boldsymbol\Psi}^\dagger(x)\sigma_z\hat{\boldsymbol\Psi}(x).
\end{equation}

This means that the properties of the system can be analyzed independently in the subspaces of a well defined number of particles appropriately distributed among the spin components. In each of these subspaces, one can find the many-body ground-state $|\mathtt{G}({\cal N},{\cal M})\rangle$ and its eigenenergy $E({\cal N},{\cal M})$. For a given number of particles $\cal N$, the global many-body ground-state of the system $|\mathtt{G}({\cal N})\rangle$ is determined as a state with the lowest energy.

The spin imbalance ${\cal M}$ is directly related to the magnetization of the system and, for convenience, we will use both names interchangeably. Note, however, that, due to the choice of units, any spin-flipping of a selected fermion changes the total magnetization of the system by $2$ quanta.

At this point, it is worth noting that the procedure of finding the ground-state of the system outlined above is correct as long as the coupling between subspaces of different spin-projections is neglected in the Hamiltonian~(\ref{Ham}). Although, from a physical point of view, this coupling is necessary for particles to flip their spins, we assume that it can be omitted. Since we consider only the ground-state and its adiabatic changes, this requirement can be fulfilled experimentally. 

\section{The Method}

The analysis of the many-body Hamiltonian (\ref{Ham}) is performed in the most natural single-particle basis built from the eigenstates $\phi_i(x)$ of the single-particle Hamiltonian (\ref{HamSingle}). For the fixed shape of the confinement $\xi$, they can be found straightforwardly by performing numerically an exact diagonalization of the operator (\ref{HamSingle}) on a dense grid in the position representation. In this approach, the single-particle Hamiltonian is represented by an appropriate tridiagonal matrix that can be easily diagonalized. As a result, one obtains shapes of the wave functions $\phi_i(x)$ and corresponding single-particle energies $E_i$ that serve as a basis for many-body calculations. The spectrum of the single-particle Hamiltonian (\ref{HamSingle}) as a function of the shape of the potential $\xi$ is shown in the right panel in Figure \ref{Fig1}. It can be clearly seen that, for $\xi=0$, single-particle energies are almost equally distributed (corresponding to an appropriate harmonic confinement), while for a large $\xi$ a quasi-degeneracy between even and odd states, characteristic for double-well confinements, emerges. 

To find the many-body ground-state of the interacting system of ${\cal N}$ particles and its energy, we numerically diagonalize the Hamiltonian (\ref{Ham}) in subspaces of fixed magnetization ${\cal M}=-{\cal N},-{\cal N}+2,\ldots,{\cal N}-2,{\cal N}$, i.e., in subspaces of given ${\cal N}_\uparrow$ and ${\cal N}_\downarrow$ values. The diagonalization is performed by representing the many-body Hamiltonian~(\ref{Ham}) as a matrix in the Fock basis spanned by vectors

\begin{equation} \label{Fock}
|i_1,i_2,\ldots i_{{\cal N}_\uparrow};j_1,j_2,\ldots j_{{\cal N}_\downarrow}\rangle = \hat{a}^\dagger_{i_1}\hat{a}^\dagger_{i_2}\cdots \hat{a}^\dagger_{i_{{\cal N}_\uparrow}}\hat{b}^\dagger_{j_1}\hat{b}^\dagger_{j_2}\cdots \hat{b}^\dagger_{j_{{\cal N}_\downarrow}}|\mathtt{vac}\rangle
\end{equation}
where $\hat{a}_{k}$ ($\hat{b}_{k}$) annihilates a fermion with spin $\uparrow$ ($\downarrow$) in the single-particle state described by the wave function $\phi_k(x)$. Ordered sets of integer numbers $\{i_1,i_2,\ldots,i_{{\cal N}_\uparrow}\}$ and $\{j_1,j_2,\ldots,j_{{\cal N}_\downarrow}\}$ determine single-particle orbitals occupied by particles with a given spin. Ordering of these sets is crucial since operators $\hat{a}_k$ and $\hat{b}_k$ fulfill fermionic anti-commutation relations, $\{\hat{a}_k,\hat{b}^\dagger_{k'}\}=\delta_{kk'}$ and $\{\hat{a}_k,\hat{a}^\dagger_{k'}\}=\{\hat{b}_k,\hat{b}^\dagger_{k'}\}=\{\hat{a}_k,\hat{b}_{k'}\}=0$.
For convenience, we assume ascending orderings: $i_1<i_2<\ldots<i_{{\cal N}_\uparrow}$ and $j_1<j_2<\ldots<j_{{\cal N}_\downarrow}$. For numerical purposes, it is not possible to consider all single-particle orbitals $\phi_i(x)$ of the single-particle Hamiltonian. Therefore, we cut $i$ on a sufficiently large integer $N_\mathrm{max}$. Numerically, we assume that the cut-off $N_\mathrm{max}$ is chosen  appropriately if the final results are not changed significantly when $N_\mathrm{max}$ is increased. Typically, for a small number of particles $N\leq 6$ and limiting interactions $g<15$, a sufficient cut-off is $N_\mathrm{max}\approx 12$. 

It is a matter of fact that the Hamiltonian (\ref{Ham}) expressed as a matrix in the basis (\ref{Fock}) belongs to the class of sparse matrices. Therefore, it is diagonalized in a straightforward manner with the Arnoldi algorithm  \cite{ArnoldiBook}. As the result of the diagonalization, one obtains the ground-state energy $E({\cal N},{\cal M})$ and the corresponding ground-state $|\mathtt{G}({\cal N},{\cal M})\rangle$ decomposed in the Fock basis (\ref{Fock}).   

The main purpose of the following analysis is to find the distribution of the particles between different spin components in the ground-state of the system, i.e., to find the ground-state magnetization~${\cal M}$. 

\section{Ground-State Magnetization}
\subsection{Harmonic Confinement}

It is quite obvious that the distribution of particles among the spin components is dependent on the interaction strength $g$, the external magnetic field $B$, and the shape of an external potential $\xi$. However, in the limit of vanishing interactions, for particles confined in an almost harmonic trap ($\xi=0$), the analysis is straightforward. In this case, whenever an external magnetic field $B$ is switched off, particles are equally distributed between spin components, i.e., each of the $N/2$ the lowest single-particle orbitals of the external confinement $\phi_i(x)$, is occupied by exactly two particles with opposite spins. Obviously, if the number of particles is odd, the highest orbital can be occupied by a particle with an arbitrary spin and therefore the many-body ground-state of the system is two-fold degenerated. When the magnetic field is relatively small, appropriate energy shifts between single-particle orbitals of opposite spins emerge and the total energy of the system increases. While the structure of the many-body state of the system with an even number of particles does not change, the degeneracy of the ground-state in the odd sector is lifted and the number of particles in each component is well~established. 

The situation described above changes when the energy shift between components, induced by the external magnetic field $B$, becomes larger than the energy difference between the last occupied and the first non-occupied single-particle orbital. Then, it is energetically favorable to flip the spin projection of the most excited fermion to decrease the energy of the ground state. In consequence, the imbalance between particles ${\cal M}$ increases. With an increasing magnetic field, the situation repeats until all particles gain the spin-flip and the system becomes fully~polarized.

\begin{figure}[!bhtp]
\centering
\includegraphics[width=\linewidth]{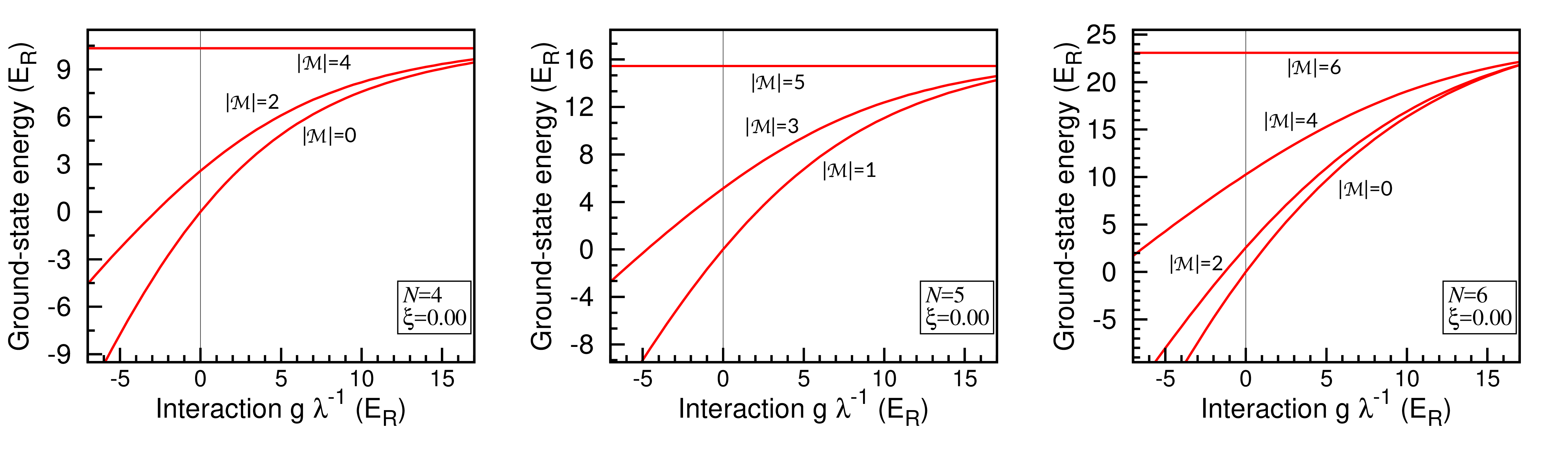}
\caption{Ground-state energies in subspaces of given magnetization $\cal M$ as functions of interaction strength $g$ in harmonic confinement $\xi=0$. The energies $E({\cal N},{\cal M})$ are always ordered according to increasing $|M|$. Therefore, the magnetization of the global ground-state for any external magnetic field $B$ can be easily determined.  \label{Fig2}} 
\end{figure}

The picture of the spontaneous magnetization of the ground-state induced by a non-vanishing magnetic field is essentially the same for an interacting system. The only difference is that now one should compare energies of interacting systems $E({\cal N},{\cal M})$ shifted by the external magnetic field. The situation is clear in the harmonic confinement, since in this case for a vanishing magnetic field and a given number of particles ${\cal N}$, the energies $E({\cal N},{\cal M})$ are ordered according to the increasing $|M|$ for any interaction $g$ (Figure~\ref{Fig2}). Therefore, there is no substantial difference between interacting and non-interacting systems when the magnetization of the ground-state is considered.  

Based on the ground-state energies for different interactions and the vanishing magnetic field, one can easily determine the phase diagram of the system when it is subjected to the external magnetic field $B$. This can be done by comparing shifted energies of the ground-states belonging to different subspaces of given magnetization, $E({\cal N},{\cal M})+B{\cal M}$. In this way, the magnetization of the many-body ground-state of the system for any $B$ and $g$ can be determined.

In the left-most column of Figure~\ref{Fig3}, we present the phase diagram of the system of a few fermions confined in a harmonic trap ($\xi=0$). As can be seen, for any interaction strength and varying magnetic field, the magnetization ${\cal M}$ is well defined. For large interactions $g$, phases with all fermions being polarized (${\cal M}=\pm{\cal N}$) are reached for arbitrarily small values of the magnetic field $|B|$. It is a quite intuitive result since, for very strong repulsions, the best way to avoid interactions is to simply polarize all fermions paying only the price for single-particle~excitations. 

\subsection{Deep Double-Well Confinement}

\begin{figure}[!bhtp]
\centering
\includegraphics[width=0.9\linewidth]{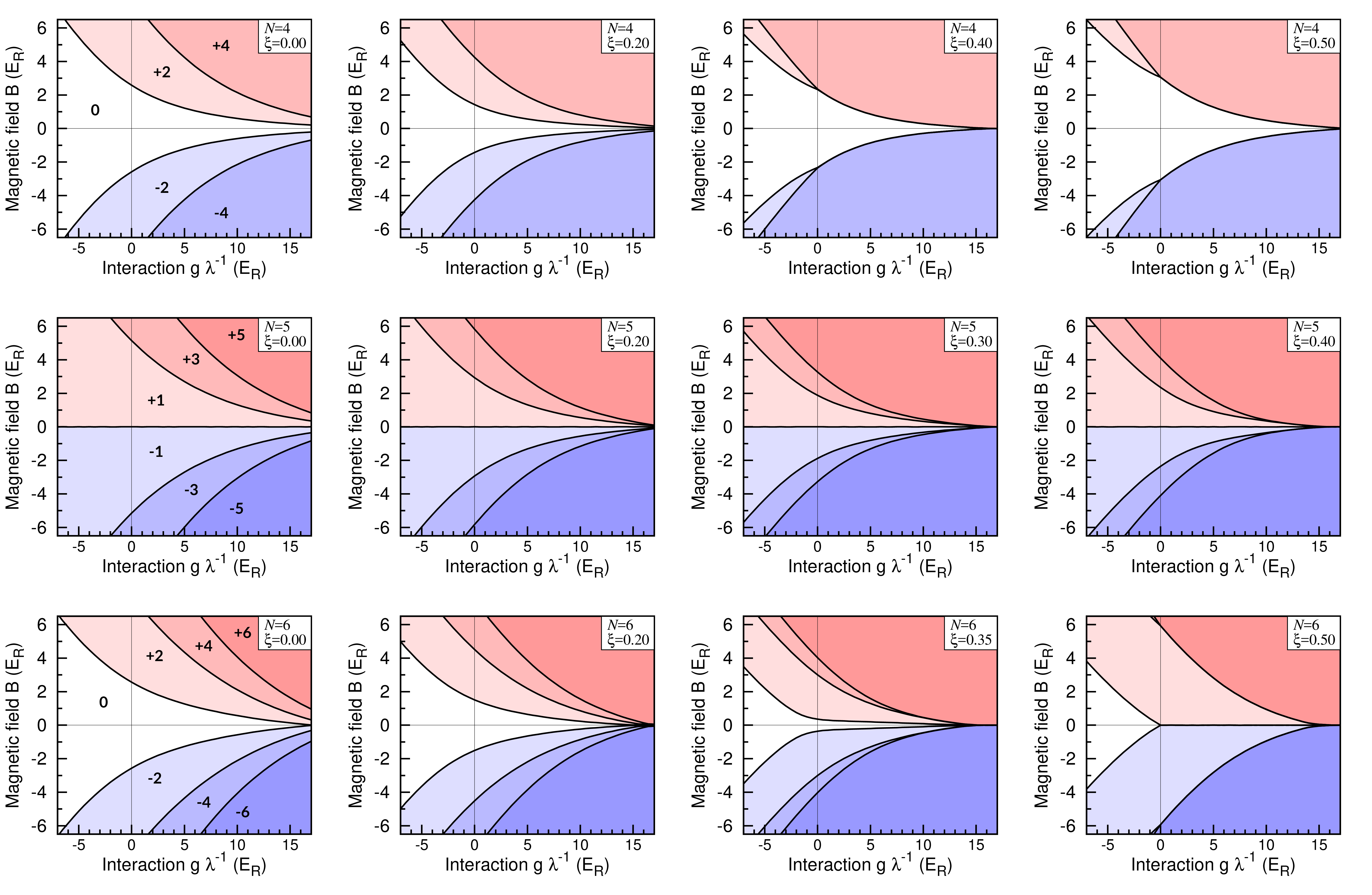}
\caption{Phase diagram of the ground-state magnetization for a different number of particles $N$ and different confinements $\xi$. For given values of a magnetic field and an inter-particle interaction, the magnetization $\cal M$ of the system is well determined (blue and red tints), and it instantly changes when on boundaries (black lines). In the limit of infinite repulsions, only the maximal magnetization can be achieved. Note that, for particular confinements (close to the deep double-well), not all magnetizations are accessible. See the main text for details.\label{Fig3}} 
\end{figure}

The situation changes significantly when the external potential is transformed to the double-well shape ($\xi>0$). In this case, if the wells are deep enough, the phase diagram of the many-body ground state changes qualitatively (see Figure~\ref{Fig3}) and for strong but finite repulsions some values of magnetization $\cal M$ cannot be reached. This observation is a direct consequence of a double-well confinement which effectively allows the system to minimize energy by spreading particles between~wells.

\begin{figure}[!hbtp]
\centering
\includegraphics[width=0.8\linewidth]{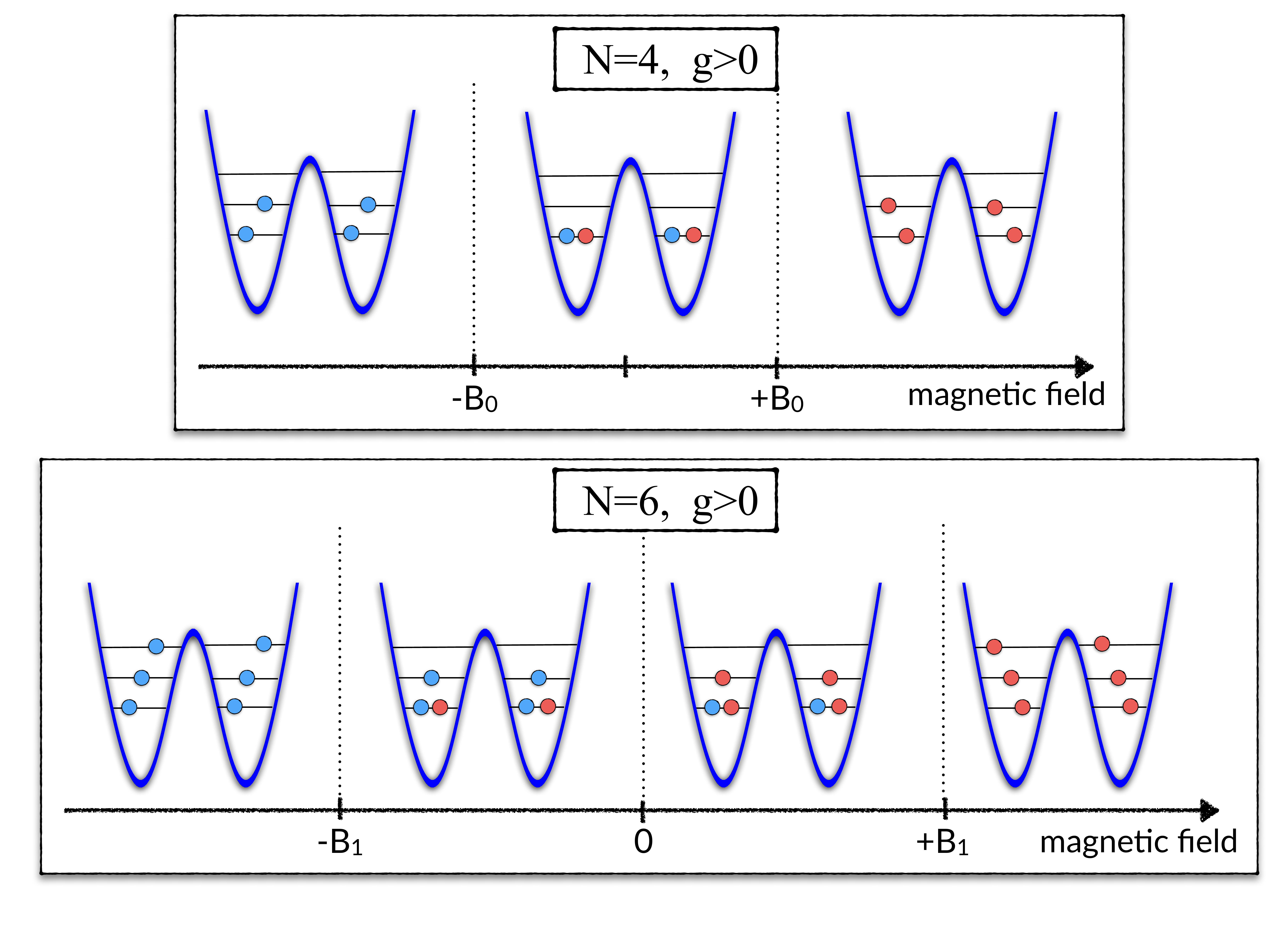}
\caption{Schematic view on a role of the external magnetic field in the magnetization of the ground-state when the system is confined in deep double-well confinement. When the magnetic field crosses some well defined critical value, appropriate particles on opposite wells simultaneously flip their spins to minimize single-particle energy. Therefore, in contrast to harmonic confinement, magnetization changes by $4$ quanta. The mechanism described for $N=4$ and $N=6$ particles can be generalized to a larger number of particles in a straightforward manner. \label{Fig4}} 
\end{figure}

To obtain a better understanding of this phenomena, let us first concentrate on the case of $N=4$ particles and deep double-well confinement $\xi=0.5$ (the right-top plot in Figure~\ref{Fig3}). When interactions are switched off ($g=0$) and the magnetic field vanishes ($B=0$) pairs of opposite-spin fermions occupy the two lowest single-particle orbitals of the single-particle Hamiltonian $\phi_0(x)$ and $\phi_1(x)$. However, due to the quasi-degeneracy of these orbitals forced by the double-well arrangement (see the right plot in Figure~\ref{Fig1}), the ground-state can also be viewed as a state in which the pairs of opposite fermions occupy single-particle orbitals localized in the left and right well, i.e., the orbitals $\varphi_{L,R}(x) = [\phi_0(x)\pm\phi_1(x)]/\sqrt{2}$. For deep enough wells, these orbitals have almost vanishing overlap of their densities $\int |\varphi_L(x)|^2|\varphi_R(x)|^2\mathrm{d}x\approx 0$, and the system behaves like a composition of two independent single-well systems with two opposite-spin particles in each well. This means that, for an increasing magnetic field, a single-particle gap emerges simultaneously in both wells. When the magnetic field is larger than a critical value $B_0$ (equal to the energy gap to the nearest excited single-particle orbitals $\phi_2(x)$ and $\phi_3(x)$), both particles flip their spins and the system becomes fully polarized (magnetization changes its value by $\Delta{\cal M}=4$, see Figure~\ref{Fig4}). A quite similar mechanism works when nonvanishing repulsions are present in the system. In this case, only the value of the critical magnetic field is different (see the top right panel in Figure~\ref{Fig3}).

In the case of $N=6$ particles, a behavior of the system in a deep double-well confinement is even more interesting. For any non-vanishing repulsion ($g>0$) and any vanishing magnetic field ($B=0$), the many-body ground-state of the system is doubly degenerated, i.e., the orbitals $\varphi_{L,R}(x)$ are occupied by pairs of opposite-spin fermions, and the remaining two particles have the same polarization (up or down) and occupy the first excited band spanned by quasi-degenerated orbitals $\phi_2(x)$ and $\phi_3(x)$. However, any non-vanishing magnetic field breaks the symmetry and the system is forced to gain appropriate magnetization ${\cal M}=\pm 2$. When the external magnetic field is strong enough (marked as $B_1$ in the bottom panel in Figure~\ref{Fig4}), two fermions from opposite wells occupying the lowest band flip their spins simultaneously, and they are promoted to the empty band spanned by orbitals $\phi_4(x)$ and $\phi_5(x)$. In consequence, the magnetization ${\cal M}$ changes its value by $4$ quanta, and the system becomes fully polarized (the bottom panel in Figure~\ref{Fig4}). For very strong repulsions, the energy gap between ground-states in subspaces of ${\cal M}=2$ and ${\cal M}=6$ vanishes and the system becomes fully polarized for an arbitrarily small magnetic field $B$, i.e., the critical value of the magnetic field $B_1$ in Figure~\ref{Fig4} approaches $0$.

These two scenarios for $4$ and $6$ particles described above can be simply generalized to other even numbers of particles. The generalization is possible since, in the limit of vanishing interactions, the higher excited band is always occupied by $2$ or $4$ fermions, and a similar picture to ${\cal N}=6$ or ${\cal N}=4$ can be addressed appropriately. This means that, as long as wells can be treated as independent, particles from opposite sites flip their spins simultaneously and magnetization changes by $4$ quanta.

The situation is quite more complicated for an odd number of particles. In this case, the unpaired particle is always delocalized between sites. Nevertheless, it is still possible to polarize the system for any interaction strength and any confinement (see the middle row in Figure~\ref{Fig3}).

\subsection{Intermediate Confinements}
For intermediate confinements, one observes a smooth transition between the two scenarios described above. When a double-well confinement is approached, regions of established magnetizations become smaller and subsequently vanish. Note that a double-well scenario appears in the system earlier for a smaller number of particles. This is a direct consequence of the structure of the single-particle spectrum of the Hamiltonian $H_\xi$ (see Figure~\ref{Fig1}). When the number of particles is larger, the higher single-particle orbitals become important. However, the characteristic quasi-degeneracy appears for larger $\xi$ (deeper double-well confinement). This fact, together with an observation that subsequent energy gaps between quasi-degeneracy manifolds decrease with increasing index $i$, leads directly to the conclusion that, along with an increasing magnetic field, particles subsequently flip their spin from the highest occupied orbitals.

\section{Conclusions}
To conclude, in this article we discussed properties of the system of a few fermions with an unconstrained spin degree of freedom. We focus on the total magnetization of the interacting system, which is established spontaneously in response to the external confinement and the magnetic field. For strong repulsions, independently of the confinement, the system remains in the state with the maximal magnetization. For intermediate interactions, whenever confinement is close to a deep double-well, only some specific values of magnetization can be achieved by the many-body ground-state. The model studied can be easily generalized to higher spin systems as well as to the bosonic particles. 
 
 
\acknowledgments{This work was supported by the (Polish) National Science Center Grant No. 2016/22/E/ST2/00555.}

\conflictsofinterest{The author declares no conflict of interest. } 
 
\reftitle{References}
\externalbibliography{yes}
\bibliography{_bibtex}
\end{document}